\documentclass[twocolumn]{svjour3}          
\usepackage[utf8]{inputenc}
\usepackage{color}
\usepackage{amssymb,amsmath}
\usepackage{graphicx}
\usepackage{authblk}
\usepackage{subfigure}
\definecolor{internationalkleinblue}{rgb}{0.0, 0.18, 0.65}
\definecolor{brickred}{rgb}{0.8, 0.25, 0.33}
\definecolor{amber}{rgb}{1.0, 0.75, 0.0}
\definecolor{applegreen}{rgb}{0.55, 0.71, 0.0}
\definecolor{darkviolet}{rgb}{0.58, 0.0, 0.83}

\newcommand{\mattia}[1]{\textcolor{brickred}{\textbf{#1}}}

\title{Multilayer control of synchronization and cascading failures in power grids}

\author{Simona Olmi \and Lucia Valentina Gambuzza \and Mattia Frasca}

\institute{Simona Olmi \at
              CNR - Consiglio Nazionale delle Ricerche - Istituto dei Sistemi Complessi, via Madonna del Piano 10, I-50019 Sesto Fiorentino, Italy,      
           \\
           INFN, Sezione di Firenze, Via Sansone 1, I-50019 Sesto Fiorentino (FI), Italy, 
           \email{simona.olmi@fi.isc.cnr.it}
           \and
           Lucia Valentina Gambuzza \at
              Department of Electrical Electronic and Computer Science Engineering, University of Catania, Catania, Italy,   \email{lucia.gambuzza@unict.it} 
              \and 
              Mattia Frasca \at
              Department of Electrical Electronic and Computer Science Engineering, University of Catania, Catania, Italy,
              \\
              Istituto di Analisi dei Sistemi ed Informatica “A. Ruberti”, Consiglio Nazionale delle Ricerche (IASI-CNR), 00185 Roma, Italy, \email{mattia.frasca@unict.it}  
}

\date{\today}

\begin{document}

\maketitle

\begin{abstract}
In this work, we propose a control scheme for power grids subject to large perturbations that cause the failure of a node of the grid. Under such circumstances, the system may lose synchrony and, in addition, a cascade of line failures can be triggered as an effect of the flow redistribution that activates the protection mechanisms equipped on each line of the grid. To devise a control action for addressing this problem, we adopt a multi-layer network-based description of the power grid that incorporates an overflow condition to model the possibility of cascading failures. The two other layers of the structure are devoted to the control, one implements the distributed proportional control law, and the other the integral control law. To exemplify the application of our model, we study the Italian high-voltage power grid for different parameters and topologies of the control layers.
\end{abstract}

\noindent \textbf{Keywords} Power grids. Swing equations. Control of complex networks. Cascading failures.

\section{Introduction}

Synchronization in power grids has become a classical example of the dynamics appearing in a system that can be modeled as a set of coupled nonlinear oscillators \cite{arenas2008synchronization,dorfler2012synchronization,witthaut2022collective}. Such modeling provides a mathematical framework that, although not capturing all the multifaceted aspects required by a very detailed description of the power system, subsumes the main characteristics of the synchronization phenomenon in a tractable way \cite{nishikawa2015comparative}. This approach, which has been pursued in many works in the field of nonlinear dynamics and control theory,  provides valuable theoretical insights on the phenomenon, allowing to characterize, for instance, the transition from the synchronous to the incoherent state (and vice-versa), stability issues, and the classes of natural frequency distributions which lead to synchronization \cite{filatrella2008analysis,choi2011complete,lozano2012role,dorfler2012synchronization,fortuna2012network,rohden2012self,carareto2013natural,motter2013spontaneous,dorfler2014synchronization,olmi2014hysteretic,grzybowski2016synchronization,olmi2016dynamics}.

Stable operation of power grids is achieved by maintaining a synchronous state in the entire network. Since power grids may be subjected to many different types of perturbations, the stability of the synchronous state is of utmost importance and many papers have investigated it. In particular, several analytical results have been obtained for network-based power grid models. For instance, a detailed stability analysis was performed in \cite{mirollo2005} and in \cite{delabays2017} for networks of classical Kuramoto oscillators (i.e., without inertia) with different topologies (namely fully coupled networks in \cite{mirollo2005} and planar graphs in \cite{delabays2017}). Regarding networks of rotators (i.e. Kuramoto oscillators with inertia), a stability analysis has been presented in \cite{nishikawa2015comparative,manik2014,coletta2016,machowski2020power}, for globally coupled networks and chain structures. In particular, the stability analysis performed in \cite{manik2014} for networks with inhomogeneous damping but identical inertia has been extended  to the case of inhomogeneous inertia and damping in \cite{coletta2016}. A  detailed stability analysis of a population of N heterogeneous rotators, randomly connected, has been reported in \cite{tumash2019stability}, showing that stable and unstable solutions can be found before stabilizing the unstable ones by a control loop.

%

The behavior of a power grid in presence of large perturbations that can eventually yield to failures that propagate along the structure is more difficult to study with analytical techniques. However, understanding cascading failures and devising mechanisms for their control is fundamental for their economic and social impact. The circumstance of cascading failures, as induced for instance by  a fault localized in a line of the grid triggering subsequent failures, has been modeled with diverse approaches. These approaches can be classified into network-based structural methods; techniques based on either DC or AC power flow calculations; or models explicitly incorporating the dynamics of the power grids. In the first approach only the structural properties of the network of interconnections are taken into account to investigate the phenomenon of cascading failures \cite{crucitti2004model,dobson2007complex,wang2014attack,zhang2016optimizing,kong2010resilience,xiao2011cascading} and devise strategies for their mitigation \cite{motter2004cascade,wang2012mitigation}. On the contrary, the second class of approaches relies on the calculation of the power flows from either DC or AC equations in order to provide a simple but tractable description of the electrical phenomena taking place in power grids  \cite{hines2010topological,soltan2015analysis,cetinay2017comparing,cetinay2018nodal,strake2019non,sturmer2021risk}. The third class of approaches makes use of models with an explicit electro-mechanical description of the dynamics of power grids. In this case, simplified models are often adopted to obtain a representation of the power grid as a system of coupled oscillators, that could enable the use of techniques and tools from nonlinear dynamics, network science and control theory \cite{demarco2001phase,yang2017cascading,schafer2018dynamically}. Although these models do not provide a very detailed description of a power system, they crucially incorporate the main characteristics of the dynamics of the generators, the loads and the mechanisms for line shut down
, as the transient dynamics can induce failures not present in the quasi-static approximation. To this purpose, either models based on a structure preserving \cite{yang2017cascading} or synchronous machine \cite{schafer2018dynamically} description of the power grid can be used. The problem of analyzing and preventing cascading failures is even more relevant in grids with high penetration of variable renewable energy sources \cite{carreras2020effects,carreras2021assessing}. 

In two recent works \cite{totz2020control,gambuzza2020distributed}, two complementary problems in control of power grids subjected to faults have been investigated. In \cite{totz2020control}, the goal of the control is to reduce the deviation from synchronization in case of faults perturbing the network dynamics. To achieve this goal, the power grid is represented as a multi-layer system \cite{boccaletti2014structure,de2021dynamics} made of two layers: the first layer represents the physical layer where the electro-mechanical phenomena of the power grid take place, while the second layer acts as control. 
On the contrary, in \cite{gambuzza2020distributed} the problem considered is the mitigation of cascading failures, induced by the dynamical evolution of the grid after a fault due to some exogenous event. The problem is addressed employing, also in this case, a two-layer representation of the power grid, where, this time, the control layer takes as input the instantaneous oscillation frequency rather than its integral. Following the formalism introduced in \cite{lombana2014distributed}, the first approach can be mapped into a layer of distributed integral controllers, while the second one into a layer of distributed proportional controllers. However, the combined use of the two control layers remains unexplored. Our paper aims at filling this gap, taking into account that proportional and integral control actions (along with the derivative mode) are widely and successfully employed in the form of the so-called PID (proportional-integral-derivative) controller since more than 70 years in many industrial applications \cite{johnson2005pid}. Although these mainly refer to controllers using a single output measurement and a single input actuator, in recent works this paradigm has been extended to distributed controllers \cite{lombana2014distributed}.

Motivated by these considerations, in this paper we investigate the combined use of the two layers independently investigated in \cite{totz2020control} and in \cite{gambuzza2020distributed} to control deviation from synchronization in presence of large perturbations leading to node removal, while simultaneously mitigating the onset of cascading failures. We adopt the multi-layer representation of the power grid, but consider three layers rather than two. The first layer represents the physical layer and is modeled with a set of swing equations (second-order Kuramoto oscillators) that also incorporate an overflow condition to take into account the intentional shut-down of a line to prevent overheating  \cite{schafer2018dynamically}. The other two layers of the multi-layer network implement proportional and integral distributed control.
Our results show that a complex interplay between the topology of the layers and the system parameters takes place, yielding scenarios where either the two layers act in a synergistic or in an antagonistic way. In addition, we find that it is difficult to derive general guidelines for the tuning of the parameters of the distributed controllers, such that this step must be accomplished by producing, for the power grid under investigation, a map of the system behavior as function of the gains of the two control layers.

The rest of the paper is organized as follows. In Sec.~\ref{sec:modelpowergrid} the model of power grid is illustrated. In Sec.~\ref{sec:results} the analysis of the Italian high-voltage power grid is discussed. In Sec.~\ref{sec:conclusions} the conclusions of the paper are drawn.

\section{Multi-layer model of the power grid}
\label{sec:modelpowergrid}

\subsection{Physical layer}

For the physical layer, we adopt a synchronous machine model \cite{nishikawa2015comparative} that incorporates an overflow condition eventually triggering cascading failures \cite{schafer2018dynamically}. According to this model, each node is associated with a rotating machine whose dynamics is described by a swing equation. Let $\mathcal{N}$ be the set of nodes with $|\mathcal{N}|=N$, $\mathcal{N}_g$ (with $|\mathcal{N}_g|=N_g$) the subset of generator nodes, and $\mathcal{E}$ (with $|\mathcal{E}|=E$) the set of links describing the lines that connect the units of the power grid. To each node $i$ with $i=1,\ldots,N$, one associates a mechanical rotor angle $\theta_i(t)$, which corresponds to the voltage phase angle, and its angular velocity $\omega_i=d\theta_i/dt$, relative to a rotating reference frame with velocity $\Omega=2\pi f$ ($f=50Hz$ or $f=60Hz$, depending on the geographical area under study). The dynamics of these variables are described by the swing equations \cite{filatrella2008analysis}:
\begin{equation}
\label{eq:swingequations}
\begin{array}{l}
\frac{d\theta_i}{dt}=\omega_i\\
I_i\frac{d \omega_i}{dt}=P_i-\gamma_i\omega_i+\sum\limits_{(i,j)\in \mathcal{E}'} K_{ij}\sin(\theta_j-\theta_i)
\end{array}
\end{equation}

\noindent where $i=1,\ldots,N$. The parameters $I_i$, $\gamma_i$, and $P_i$ represent the rotating machine inertia, damping coefficient, and power. The power can be either positive, $P_i>0$, for nodes that act as generators, injecting the power into the system, or negative, $P_i<0$, for nodes that act as loads, absorbing the power from the system. Here $\mathcal{E}' \subseteq \mathcal{E}$ represents the set of the operating (i.e., not failed) links of the power grid. The parameters $K_{ij}$ are the elements of the weighted adjacency matrix describing its topology, and are related to the electrical quantities characterizing the nodes by the relationship $K_{ij}=B_{ij}V_iV_j$ where $B_{ij}$ is the susceptance between nodes $i$ and $j$, and $V_i$ and $V_j$ are the voltage amplitudes. Eqs.~(\ref{eq:swingequations}) hold under several assumptions that allow to simplify the power flows equations governing the electrical system. In particular, the voltage amplitudes $V_i$ are assumed to be constant, the ohmic losses negligible, and the variations in the angular velocities, $\omega_i$, small compared to the reference $\Omega$.

Important quantities to determine the occurrence of failures due to line overloads are the flows that are defined as follows:
\begin{equation}
\label{eq:flowdefinition}
F_{ij}(t)=K_{ij}\sin(\theta_j(t)-\theta_i(t))
\end{equation}

\noindent $\forall (i,j) \in \mathcal{E}$. The maximum flow that a line can accommodate is $F_{ij}=K_{ij}$. Since ohmic losses induce overheating in the lines, connections are shut down when the flow exceeds a fraction $\alpha \in [0,1]$ of its maximum, which corresponds to set the line capacity as $C_{ij}=\alpha K_{ij}$, where $\alpha$ is a tunable parameter of the model. Hence, the overload condition for a generic line $(i,j)$ is given by:
\begin{equation}
\label{eq:overloadcondition}
|F_{ij}(t)|>C_{ij}=\alpha K_{ij}
\end{equation}

When this condition is met at some time, then the line is shut down. This results in a change of the topology that elicits a further dynamical redistribution of the flows and eventually triggers other line protection mechanisms, yielding a cascading failure. The purpose of the control layers, described in the next sections, is to prevent these failures, while maintaining synchronization.

\subsection{Control layers}

\begin{figure}
\centering
\includegraphics[width=0.45\textwidth]{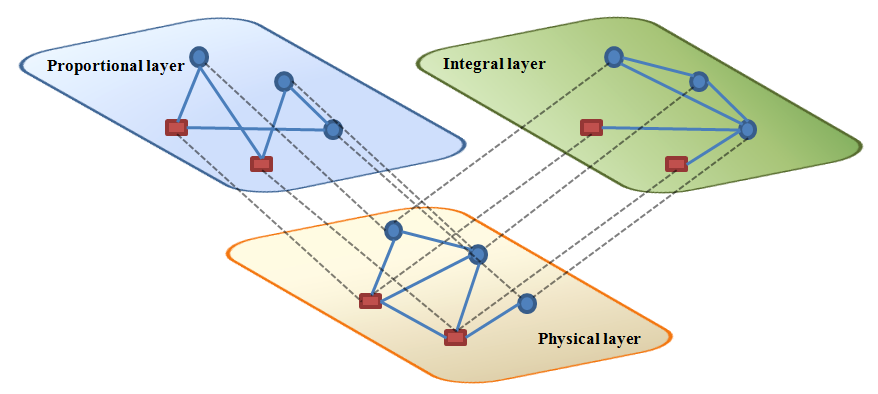}
\caption{Multi-layer model of the power grid with control as in Eq.~\eqref{eq:controlterm}. The bottom layer represents the physical layer, nodes are generators/loads of the power system and intra-layer links are the lines of the power system. The other layers represent distributed controllers implementing the proportional control law \eqref{eq:controltermProp} and the integral control law \eqref{eq:controltermInt}, respectively. The inter-layer links allows the control law to be applied to the controlled node of the physical layer. \label{fig:multilayerstructure}}
\end{figure}

In our multi-layer structure, control is implemented in two layers whose nodes are in one-to-one corrispondence with the units of the physical layer. The overall structure is thus composed of three layers, as schematically shown in Fig.~\ref{fig:multilayerstructure}. The control input can be viewed as the signal associated to the inter-layer link between corresponding nodes of the multi-layer networks. Intra-layer links of the physical layer, instead, represent the lines of the grid (that is, the channels through which energy exchange between two power units occurs). Finally, the intra-layer links of each control layers represent flows of information to build the control law.

In the presence of the two control layer, the equations describing the dynamics of the power units can be written as \cite{totz2020control,frasca2021control}:

\begin{equation}
\label{eq:swingequationscontrolled}
\begin{array}{lll}
\frac{d\theta_i}{dt} & = &\omega_i\\
I_i\frac{d \omega_i}{dt} & = & P_i-\gamma_i\omega_i+\sum\limits_{(i,j)\in \mathcal{E}'} K_{ij}\sin(\theta_j-\theta_i)+u_i
\end{array}
\end{equation}

The term $u_i(t)$ represents the control input, and is here obtained as the sum of two contributions:

\begin{equation}
\label{eq:controlterm}
u_i=u_i^P+u_i^I
\end{equation}

The term $u_i^P(t)$ is a distributed action obtained by setting for each link a control law proportional to the difference of the frequencies at the extremes \cite{frasca2021control}:

\begin{equation}
\label{eq:controltermProp}
u_i^P=G_P\xi_i^P\sum\limits_{j=1}^N a_{ij}^p(\omega_j-\omega_i)
\end{equation}

The term $u_i^I$ is also a distributed action, that, however, implements an integral action, defined by specifying the dynamics of the term itself \cite{totz2020control}: 

\begin{equation}
\label{eq:controltermInt}
\dot{u}_i^I=G_I\xi_i^I\sum\limits_{j=1}^N a_{ij}^I(\omega_j-\omega_i)
\end{equation}

In both Eq.~\eqref{eq:controltermProp} and~\eqref{eq:controltermInt}, $\xi_i^P$ and $\xi_i^I$ are binary variables, allowing to select which units of the power grid are subject to control, that is, $\xi_i^P=1$ ($\xi_i^I=1$) if node $i$ is controlled by a proportional (integral) control action, and  $\xi_i^P=0$ ($\xi_i^I=0$) otherwise. With the term pinning control we refer to the case when not all nodes are controlled, and we call pinned nodes the nodes for which $\xi_i^h=1$ with $h=\{P,I\}$. $G_P$ and $G_I$ represent the gains of the two layers, and $a_{ij}^P$ and $a_{ij}^I$ are the coefficients of the adjacency matrices $A^P=\{a_{ij}^P\}$ and $A^I=\{a_{ij}^I\}$ encoding the topologies of the two layers.

In summary, the dynamics of the power grid with the control law \eqref{eq:controlterm} is described by:

\begin{equation}
\label{eq:swingequationscontrolled}
\begin{array}{lll}
\frac{d\theta_i}{dt} & = &\omega_i\\
I_i\frac{d \omega_i}{dt} & = & P_i-\gamma_i\omega_i+\sum\limits_{(i,j)\in \mathcal{E}'} K_{ij}\sin(\theta_j-\theta_i)\\
& & +G_P\xi_i^P\sum\limits_{j=1}^N a_{ij}^P(\omega_j-\omega_i)+G_I\xi_i^I\sum\limits_{j=1}^N a_{ij}^I(\theta_j-\theta_i)
\end{array}
\end{equation}

\noindent where $i=1,\ldots,N$. Here we note that the control signal can be interpreted as power injection for positive $u_i$ or power absorption for negative values of $u_i$, which, for loads, can be obtained by modulating the effective power associated to the bus and, for generators, can be realized using storage devices (e.g., batteries) that can absorb or inject power to the bus \cite{qian2010}.

\subsection{Case study and measures characterizing the system behavior}

We focus our analysis on the case study of the Italian high-voltage (380kV) power grid \cite{GENI,rosato2007topological,filatrella2008analysis,fortuna2012network,tumash2019stability}. Available data on the structure of this power grid are used to set the topological characteristics of the physical layer. The network is assumed to be homogeous and undirected, that is, $K_{ij}=K_{ji}=Ka_{ij}$, where $a_{ij}$ are the coefficients of the adjacency matrix modeling the lines of the power grid, i.e., $a_{ij}=1$ if the power units $i$ and $j$ are connected by a power line, and  $a_{ij}=0$ otherwise. The network contains $N=127$ nodes ($34$ generators and $93$ loads) and $L=171$ links. The parameters of the model have been selected based on previous works \cite{totz2020control,schafer2018dynamically} and in order to meet the condition that the network is synchronized in the absence of faults. In more details, we set as $I_i=I=10$ $\forall i$, $\gamma_i=\gamma=1$ $\forall i$, $K=11$, $P_i=-1$ for the load nodes, and $P_i=2.735$ for the generation nodes as in~\cite{totz2020control} such that the network is balanced, i.e., $\sum\limits_{i=1}^N P_i=0$. This condition is a pre-requisite for synchronization. Finally, the parameter $\alpha$ has been set as $\alpha=0.8$ \cite{schafer2018dynamically}.

For the other two layers used in Eqs.~\eqref{eq:swingequationscontrolled} we tested the effect of several topologies, as described in the following sections. 

In order to monitor the behavior of the multi-layer model of the power grid, we use several indicators to measure synchronization and the number of failed lines. First of all, we consider the Kuramoto order parameter $R(t)$ defined as follows:

\begin{equation}
R(t)e^{\mathrm{i} \Phi (t)}=\frac{1}{N} \sum\limits_{j=1}^{N} e^{\mathrm{i} \theta_j(t)}
\end{equation}

This parameter takes values in $[0,1]$, with values close to one indicating that the phases are synchronized, and values close to zero denoting the absence of phase synchronization. Likewise, it is important to monitor not only the level of phase synchronization, but also that of frequency synchronization. To this aim, we consider the standard deviation of the instantaneous frequencies:

\begin{equation}
\label{eq:deltaomega}
\Delta \omega (t)= \sqrt{\frac{1}{N}\sum\limits_{j=1}^N(\omega_j(t)-\bar{\omega}(t))^2}
\end{equation}

\noindent where $\bar{\omega}(t)=\frac{1}{N}\sum\limits_{i=1}^N \omega_j(t)$ is the instantaneous average frequency of the power units. The parameter $\Delta \omega (t)$ provides information about the deviation from complete frequency synchronization, with values close to zero indicating that all nodes of the grid oscillate at the same frequency. Alongside with this parameter, we also monitor the value of the frequencies $\omega_j$, with $j=1,\ldots,N$, as under normal operation of the power grid these quantities must be zero or very close to zero.

To provide a more comprehensive understanding of failures under different control gains we also report the power loss $P$ of all loads, which  represents the difference between the initial total power of all load nodes and the total effective power of all load nodes at time $t>0$:

\begin{equation}
  P=  \frac{1}{\mathcal{N}_l}\sum_{i}^{\mathcal{N}_l}(P_i+u_i)- \frac{1}{\mathcal{N}_l}\sum_{i}^{\mathcal{N}_l}P_i ,
\end{equation}
where $\mathcal{N}_l$ is the total number of loads.

Finally, to monitor the lines that are not operative in the system, or, alternatively, those that are active, it is useful to consider two different measures. We indicate with $n_c$ the number of lines failed during the window of time where Eqs.~\eqref{eq:swingequationscontrolled} are simulated. In more detail, for each line of the power grid, we calculate the flow in the line and check condition~\eqref{eq:overloadcondition} at each time $t$. If, at some time, the flow exceeds the maximum capacity, then the line is shut down for the rest of the simulation and, hence, is considered in the count of failed lines. However, in our analysis we will start from a fault located in one of the power grid units and assume that this fault removes the nodes and all its lines from the power system. The lines removed in this way are not shut down because of an overflow and are, therefore, not counted in $n_c$. To take into account also the failures of these lines, when appropriate, we will consider the number of active links.

\subsection{Topology of the layers}
\label{integral_topologies}


For the sake of clarity, we focus our analysis on the Italian high-voltage power grid, even if our approach can be extended to other physical layer topologies. Instead, we emphasize different control layer topologies, as they play a crucial role in our investigation. The distributed nature of the controllers, which gather information from neighboring units to determine the appropriate control action at each node, make this aspect particularly significant.

For what concerns the proportional layer, we have considered that all nodes are controlled, namely $\xi^P_i=1$, $\forall i$. For the layer topology, which rules how the proportional controllers are connected each other (or, equivalently, which information are available at each node of the layer), we have analysed a connectivity identical to that of the physical layer and some random networks, that for simplicity we have generated using the Erd\"os-R\'enyi (ER) model \cite{latora2017complex}. As discussed in Sec.~\ref{sec:results}, for the control layer we first carry out a preliminar analysis to check whether the layer, in the absence of the integral control, can prevent cascading failures triggered by failures in any of the power system units.

In the case of integral control, only the generator nodes are controlled, that is, $\xi^I_i=1$ for $i\in \mathcal{N}_g$. The topology of this layer is obtained starting from an existing network in two different ways. Let us consider the adjacency matrix $\bar{A}$ of a given graph, whose nodes are the units of the power grid. From this network, we extract the subgraph where the nodes are all the generators of the power grid and their first neighbors, and the edges are the connections among these nodes. We name this network as the local network extracted from $\bar{A}$ and indicate its adjacency matrix as $A^{loc}(\bar{A})$. The elements of this matrix are given by:

\begin{align} \label{cij_loc_def}
a^{loc}_{ij}(\bar{A})= 
\begin{cases}
\bar{a}_{ij} \,\text{,} & \text{if}~ i \vee j \in \mathcal{N}_g \\
0 \,\text{,} &\text{otherwise}
\end{cases}
\,.
\end{align}

In the second case, in addition to the links obtained in this way, we also consider each possible connection between any pair of generators. The network obtained in this way, that we call extended topology, has adjacency matrix indicated as $A^{ext}(\bar{A})$, whose elements are given by:

\begin{align} \label{cij_ext_def}
a^{ext}_{ij}(\bar{A})= 
\begin{cases}
1 \,\text{,} &i\wedge j\in \mathcal{N}_{g}\\
a^{loc}_{ij}(\bar{A}) \,\text{,} &\text{otherwise}
\end{cases}
\,.
\end{align}

\begin{figure*}[ht]
		\includegraphics[width=0.32\linewidth]{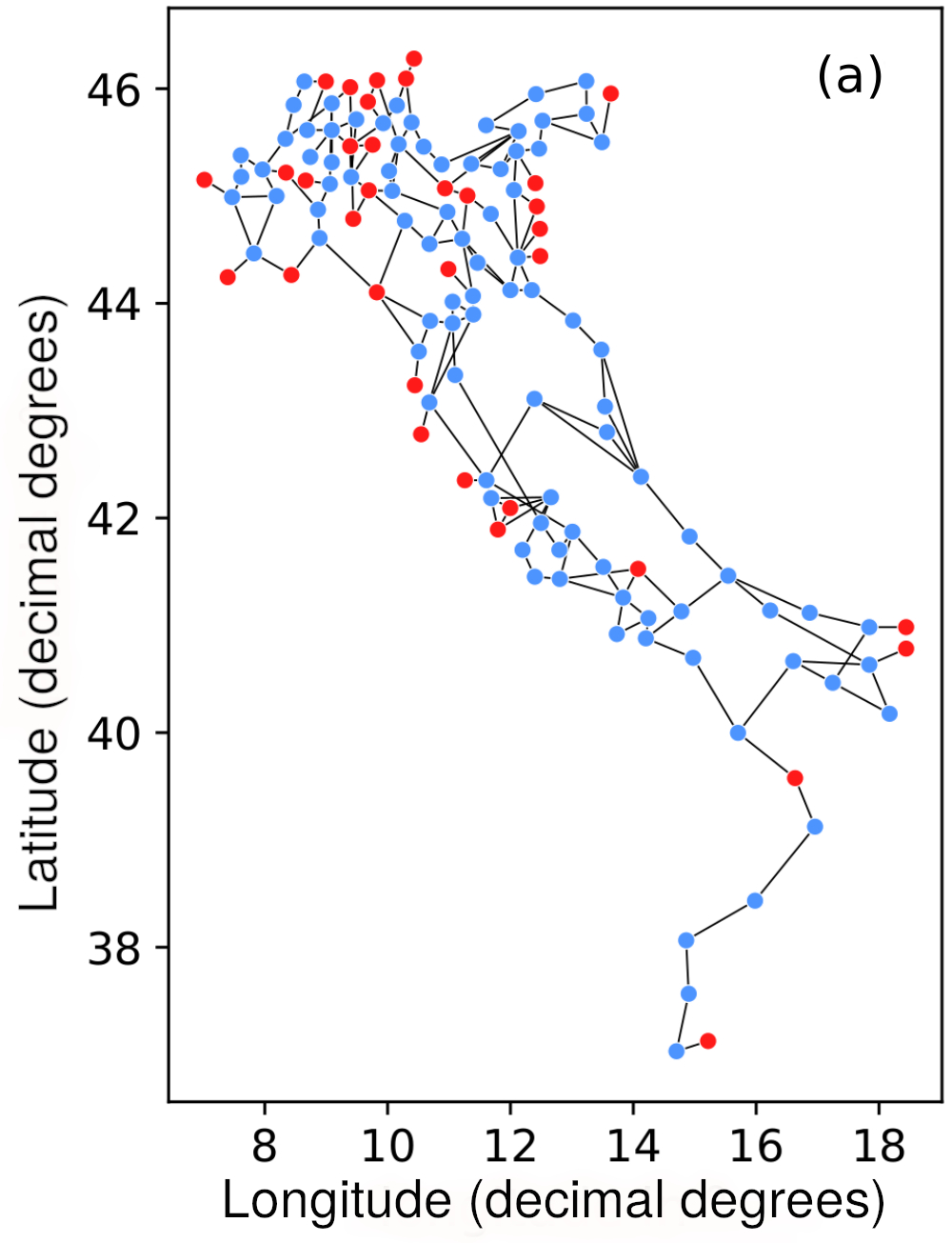}
		\includegraphics[width=0.32\linewidth]{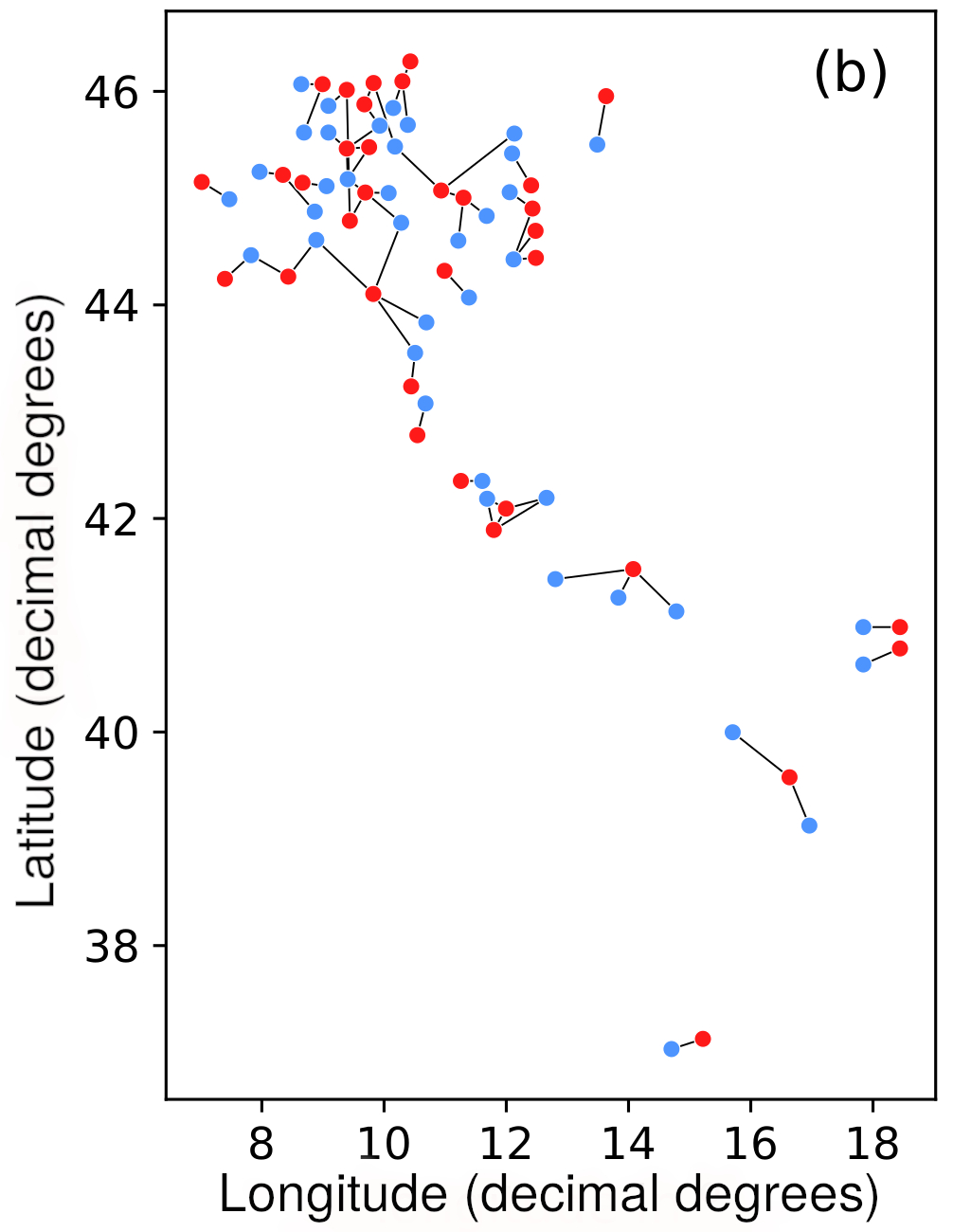}
		\includegraphics[width=0.32\linewidth]{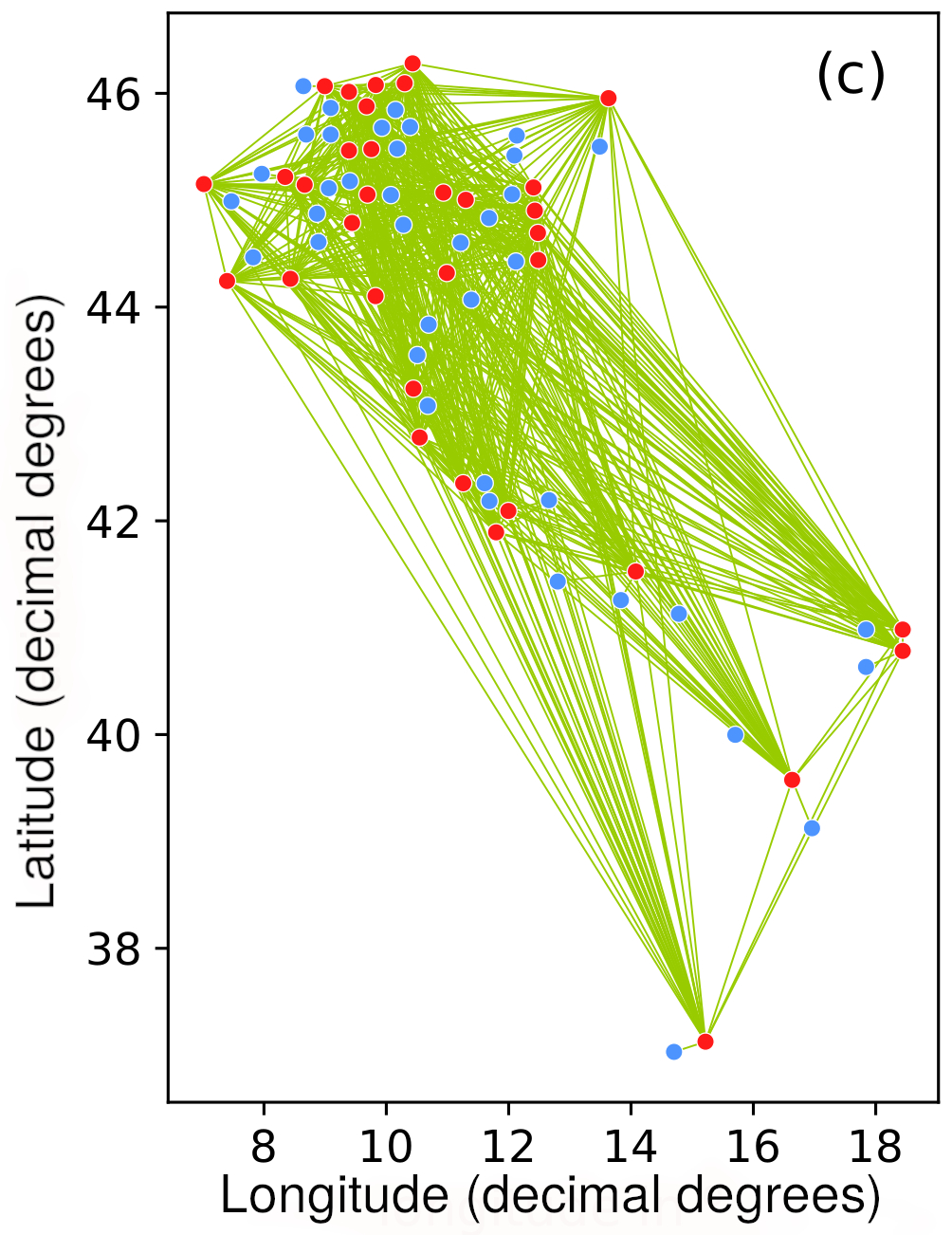}
	\caption{Visualisation of the topology of individual layers of the multi-layer power grid:
		(a) Topology $a_{ij}$ of the physical layer based on the real Italian high-voltage power grid.
		(b) Integral control layer topology $A^I=A^{loc}(A)$ where the communication links of the generators are as in the physical layer.
		(c) Integral control layer topology $A^I=A^{ext}(A)$ where generators possess additional communication links to all other generators in the network (green). Red nodes denote generators, while blue nodes denote consumers. Position of nodes has been slightly modified to improve readability.
	\label{fig:topologies}
 }
\end{figure*}

In the following, for the integral layer we will analyse local topologies obtained from either the physical or the proportional layer connectivity, namely $A^I=A^{loc}(A)$ or $A^I=A^{loc}(A^P)$ as well as the extended topology obtained from the physical layer connectivity, namely $A^I=A^{ext}(A)$. For the sake of illustration, in Fig.~\ref{fig:topologies}(b) we show the local topology obtained from the Italian high-voltage power grid, with adjacency matrix $A^I=A^{loc}(A)$, while in Fig.~\ref{fig:topologies}(c) we show the extended topology, with adjacency matrix $A^I=A^{ext}(A)$.


\section{Results}
\label{sec:results}

\subsection{Control of cascading failures}
\label{cascading_failures}

In this section we study the control of cascading failures that can be triggered by a fault in one node of the grid when only the proportional layer is active, i.e., $G_I=0$ in Eqs.~\eqref{eq:swingequationscontrolled}. Although the structure of the control layer is similar to the one considered in \cite{frasca2021control}, the problem here investigated is more general and, often, more complicated, due to the fact that the initial fault is located in a node of the grid, rather than in a line. Indeed, we will show that its solution requires the use of a topology for the proportional control layer different than that of the physical layer.

Preliminary to the analysis of the effect of the proportional control layer, we have studied the model in Eqs.~\eqref{eq:swingequationscontrolled} in the absence of any control, i.e., $G_P=G_I=0$. The topology of the physical layer is given by the Italian high-voltage power grid described in Sec.~\ref{sec:modelpowergrid}. The same parameters are used for all generators and loads, i.e., $\gamma_i = \gamma$, $\forall i$ (with $\gamma=0.1$), $I_i = I$ (with $I = 1$). In addition, we set the same susceptance for all edges, i.e., $K_{i,j} = K$ ($K=11$). For each node of the power grid, we have considered a fault located in the node, by removing the node and all its links, and simulated Eqs.~\eqref{eq:swingequationscontrolled}, using condition  \eqref{eq:overloadcondition} to check if at some time there are lines that fail. We have then counted the number of failed nodes, $n_c$, and named as \emph{critical} those nodes for which $n_c \neq 0$. We have found that, under these conditions, there are 24 critical nodes in the power grid.

\begin{figure}
\includegraphics[width=0.45\textwidth]{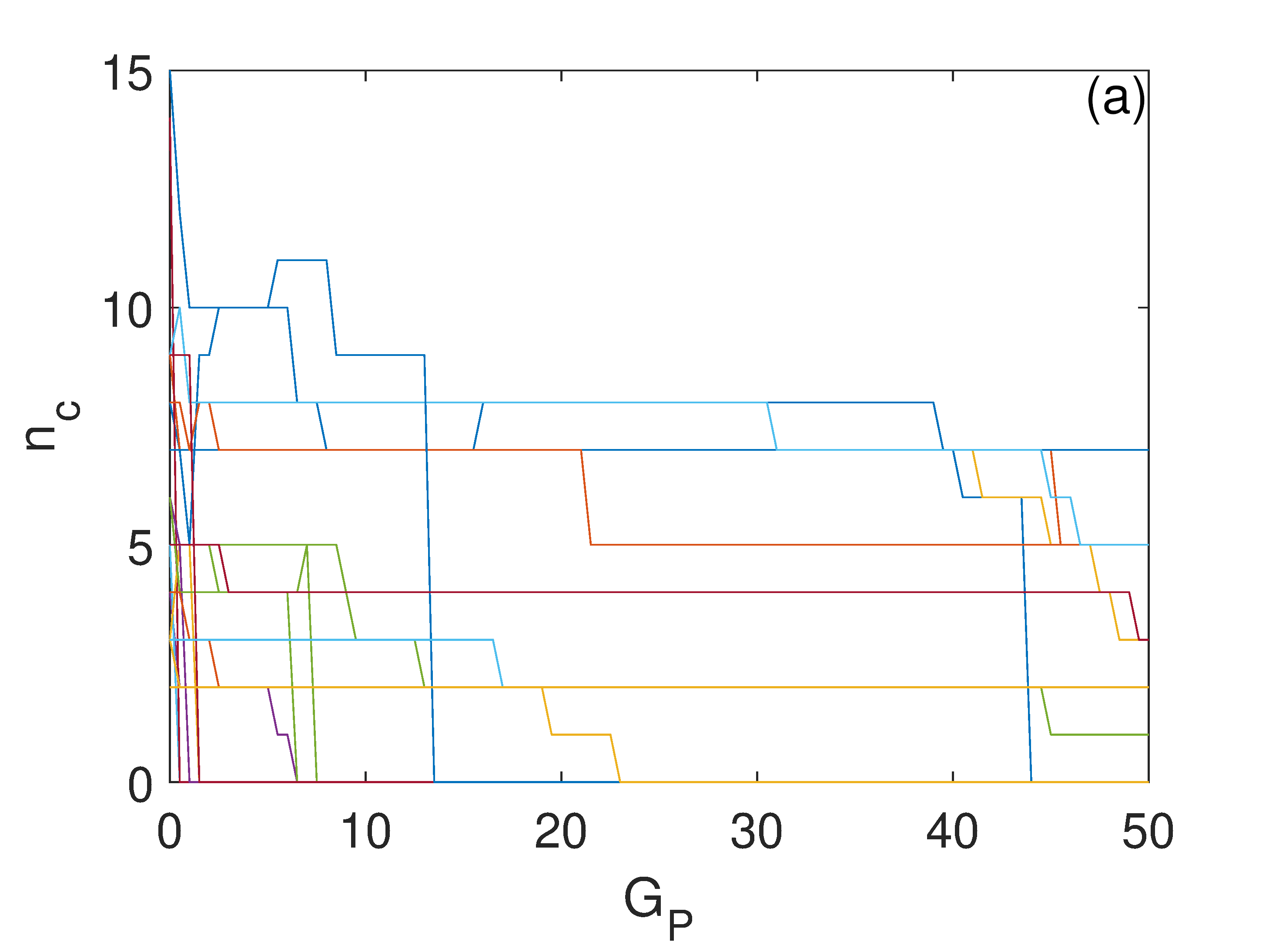}
\includegraphics[width=0.45\textwidth]{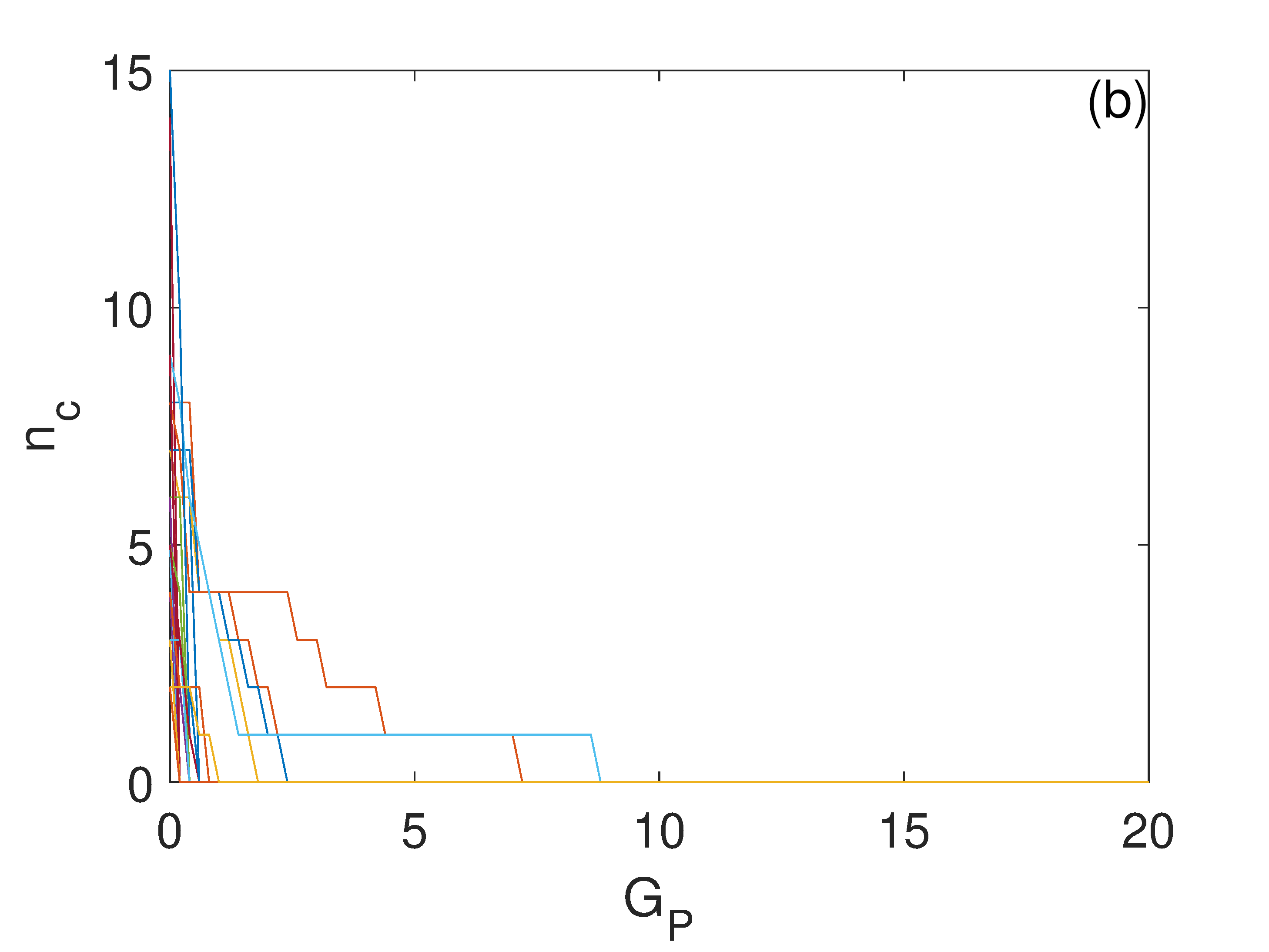}
\caption{Control of cascading failure in the Italian high-voltage power grid by the proportional layer. The different curves show the number of faulty lines, $n_c$, as a function of the coupling gain $G_P$ when the initial fault is located in one of 24 critical nodes of the power grid. Here, the control layer topology is given by (a) the same network as in the physical layer, and (b) an ER network with $p=0.04$.\label{fig:SoloProp}}
\end{figure}

In \cite{frasca2021control} it is shown that failures triggered by an initial fault in one of the grid lines can be controlled by a proportional layer having the same connectivity of the physical one. Motivated by these results, as a first case study here we have considered the same assumption for the topology of the control layer. We have thus carried out numerical simulations of Eqs. \eqref{eq:swingequationscontrolled} for $\xi^P=1$ $\forall i=1,\ldots,N$ and different values of $G_P$. For each of the twentyfour critical nodes of the power grid, determined with the previous analysis, we have calculated $n_c$ as a function of the gain $G_P$ of the control layer, when the initial fault is located in the considered node. The results are reported in Fig.~\ref{fig:SoloProp}a, which shows that there are thirteen nodes for which the cascading failure cannot be controlled, even when using a large value of $G_P$. Hence, although for many nodes this approach is effective, it is not for all nodes of the grid. To investigate whether this is due to the control law itself or to the topology of the control layer, we have then repeated the analysis for different configurations of the control layer. In particular, we have considered a control layer where the links among nodes are produced by the ER model, with each pair of nodes being connected with probability $p$. We have found that, even at low values of $p$, this approach is effective to control cascading failures triggered at any node of the power system. Fig.~\ref{fig:SoloProp}b shows the result for a network obtained with $p=0.04$, for which cascading failures are prevented for all critical nodes by setting a large enough value of $G_P$. Similar results are found for other ER networks with the same value of $p$. The value of $p$ itself does not appear to be a critical parameter, as long as it ensures a sufficient connectivity among the nodes.


\subsection{Control of synchronization and cascading failures}

As shown in the previous section, in the presence of the proportional layer only, a larger number of cascading failures can be controlled either by increasing $G_P$ or choosing a proper control topology, different from the physical one. This analysis does not consider the level of synchronization, as for instance measured by the parameter $\Delta \omega$.  In order to keep a high level of network synchronization while controlling the cascading failures, we resort to an additional control strategy, corresponding to an additional control layer. However, the response of a complex system to a node perturbation is a non-trivial phenomenon due to the interplay of many factors that concur to determine it, such as the typology and duration of the perturbation, the weight of the edges (in this case representing the susceptance of the power lines), the connectivity of the perturbed node, the topology of the physical layer, the topology of the control layers, and many others. This has two consequences. On the one hand, the inclusion of another layer, and in particular a dynamical one as the integral layer, makes possible to observe cascading failures also initiated by those nodes not classified as critical in the absence of the integral layer. On the other hand, the interplay among all these factors makes extremely complicated to exhaustively investigate the role played by each of them. For this reason, in this manuscript we limit ourselves to exemplifying some typical scenarios that are observed, considering different topologies for the control layers. Regarding  the topology of the proportional control layer, we consider two cases: i) an ER network generated with connection probability $p=0.04$; ii) the Italian high-voltage power grid. Regarding the integral control layer, we consider three different topologies that are built either from the network backbone of the Italian power grid or from the ER network generated with connection probability $p=0.04$, as detailed in Sec. \ref{integral_topologies}.

\begin{figure*}
\includegraphics[width=1\textwidth]{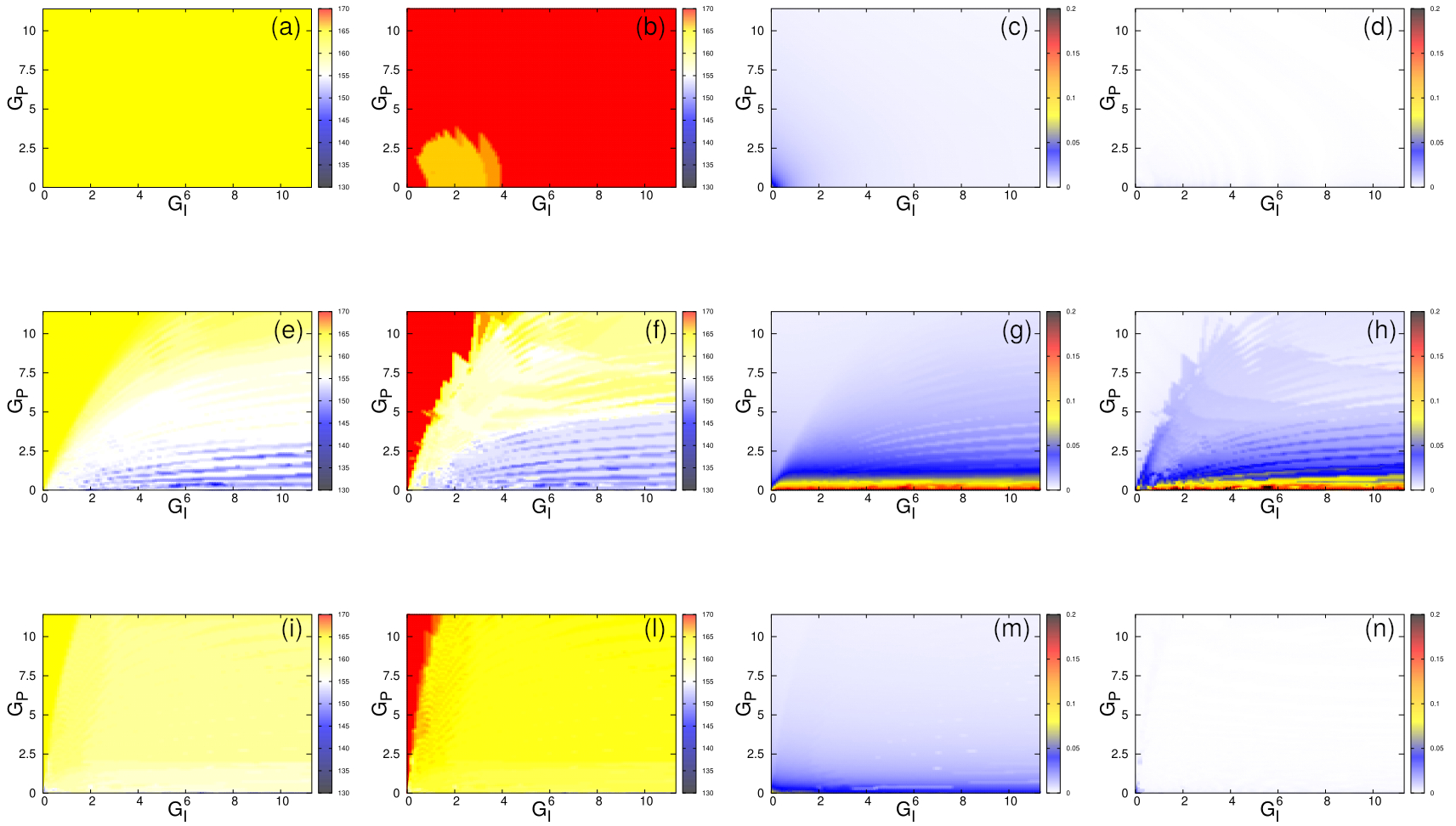}
\caption{Single node perturbation when the proportional layer is given by an ER network with $p=0.04$: removal of node 24. First two columns: active links. Third and fourth column: standard deviation $\Delta \omega$ for different values of $G_P$, $G_I$. (a)-(d): Global control topology in the integral layer, i.e., $A^I=A^{ext}(A)$, when the perturbation is active (panels a and c) and when the perturbation is ended (panels b and d). (e)-(h): Local topology in the integral layer, i.e, $A^I=A^{loc}(A)$, when the perturbation is active (panels e and g) and when the perturbation is ended (panels f and h).  (i)-(n): Local control topology in the integral layer, i.e., $A^I=A^{loc}(A^P)$, when the perturbation is active (panels i and m) and when the perturbation is ended (panels l and n). Parameters: $I_i=I=10$ $\forall i$, $\gamma_i=\gamma=1$ $\forall i$, $\alpha=0.8$, $K=11$, $P_i=-1$ for the load nodes, and $P_i=2.735$ for the generation nodes. \label{fig:nodo24extended}}
\end{figure*}

We start our analysis by considering the dynamics emergent in the system when the initial fault (causing node removal along with its edges) is located in node 24. Node 24 is a generator connected to three other units, and its removal causes the failure of only these three lines. We have investigated, for a wide range of the control parameter values, the effectiveness of the control schemes, varying both the topology of the proportional control layer and the integral control layer. In particular in Fig.~\ref{fig:nodo24extended}, we have varied the integral control layer topology, while keeping fixed the proportional layer topology, chosen to be the ER network with $p=0.04$ introduced in Sec.~\ref{cascading_failures}. The figure illustrates the system behavior when the node is first removed from the grid and then reconnected to it at the end of the perturbation, as a transient dynamics eventually leading to cascading failures can be elicited either by the removal or reactivation of the node.

Using the extended control topology $A^I=A^{ext}(A)$ for the integral control layer, our approach is able to fully prevent cascading failures occurring when the node is removed from the grid (panel a). When the node is reconnected to the network (panel b), the control works almost everywhere in the parameter regions considered. Moreover, this control guarantees a high level of synchronization, as the standard deviation $\Delta \omega$ remains almost zero both during the perturbation (panel c) and when the perturbation stops and the node is reconnected to the grid (panel d). 

When the local topology, i.e., $A^I=A^{loc}(A)$ is used in the integral control layer, when the perturbation is effective (panel e) we observe cascading failures for small values of $G_P$ and large values of $G_I$. They can be successfully controlled using large values of $G_P$ and small values of $G_I$. This is also evident in the behavior observed after reconnecting the perturbed node to the grid (panel f), that shows how the red region corresponding to full control of cascading failures shrinks to the upper left corner of figure.
The level of synchronization is poorer than in the previous case, both during the perturbation (panel g) and when the perturbation is over (panel h). In particular, lower levels of synchronization are observed for smaller values of $G_P$. 

Finally, if the topology for the integral control layer is that induced by the proportional layer, i.e., $A^I=A^{loc}(A^P)$, we observe similar effects with respect to mitigation of cascading failures (panel l is comparable to panel f even if the red region shrinks in panel l), while synchronization control is more effective, even at small values of $G_P$ (see panels m,n).

\begin{figure}
\centering
\includegraphics[width=0.45\textwidth]{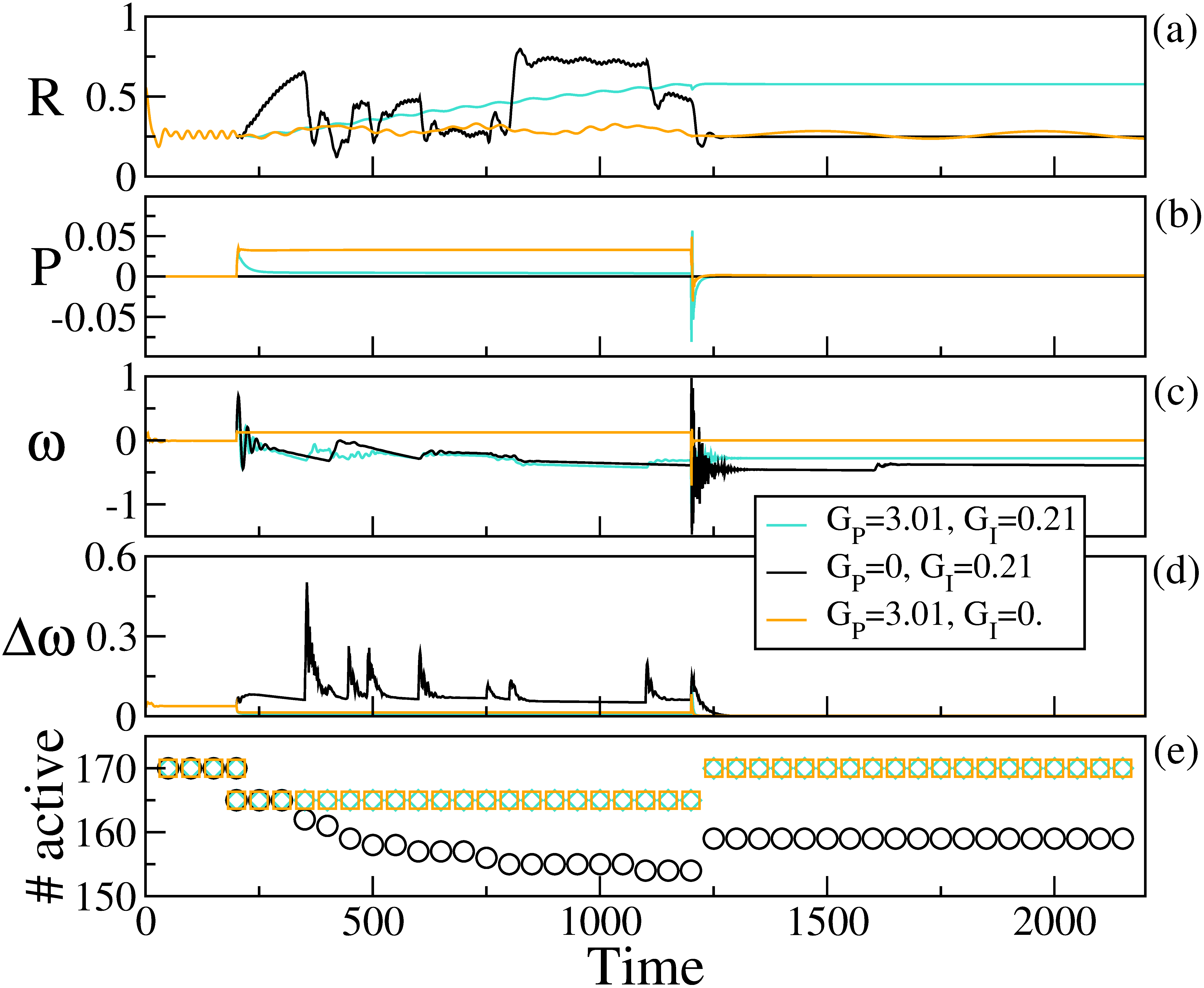}
\caption{Time behavior of the order parameter $R$ (a), the power loss of all loads (b), the frequency $\omega_i$ of the perturbed node (c), the standard deviation $\Delta \omega$ (d), and the number of active links (e) when node 24 is removed. The proportional layer is an ER random network with connection probability $p=0.04$. The integral layer is characterized by the adjacency matrix $A^I=A^{loc}(A^P)$. The other parameters are set as in Fig. \ref{fig:nodo24extended}. \label{fig:nodo24timebehavior}}
\end{figure}

To better illustrate the role played by each control layer separately, we now focus on the time evolution of the system at selected values of the control gains $G_P$ and $G_I$. In Fig.~\ref{fig:nodo24timebehavior} we show the time evolution of one of the best cases, according to Fig.~\ref{fig:nodo24extended} (panels l and n), and compare it to the cases when only one mode of the controllers is active. The perturbation is applied in the time interval $t\in [200, 1200]$, during which the perturbed node is disconnected in the physical layer. When both controls are present (light blue curves), cascading failures are prevented and, once the node is reconnected, all links are active (panel e). At the same time, $\Delta \omega\approx 0$ during all the simulation time interval (panel d), while, when the perturbation ends, the Kuramoto order parameter reaches values of $R$ ($R\approx 0.6$) slightly higher than those before the perturbation (panel a). If only the integral control is applied (black curves), the level of synchronization in the network is higher, on average, during the perturbation (panel a), but we observe a large number of failed lines during the perturbation ($n_c=15$) that prevents the system from recovering when the perturbation ends (panel e). Finally, when the integral control is turned off, while keeping the proportional control on (orange curves), we obtain a low level of synchronization ($R\approx 0.24$ in panel a) during all the time window and we observe a larger power loss (panel b), when the perturbation is active, with respect to the other cases. Finally, it is evident the role played by the proportional control in preventing the cascading failures, since all lines are active when the perturbation ends (panel e).

\begin{figure*}
\includegraphics[width=1\textwidth]{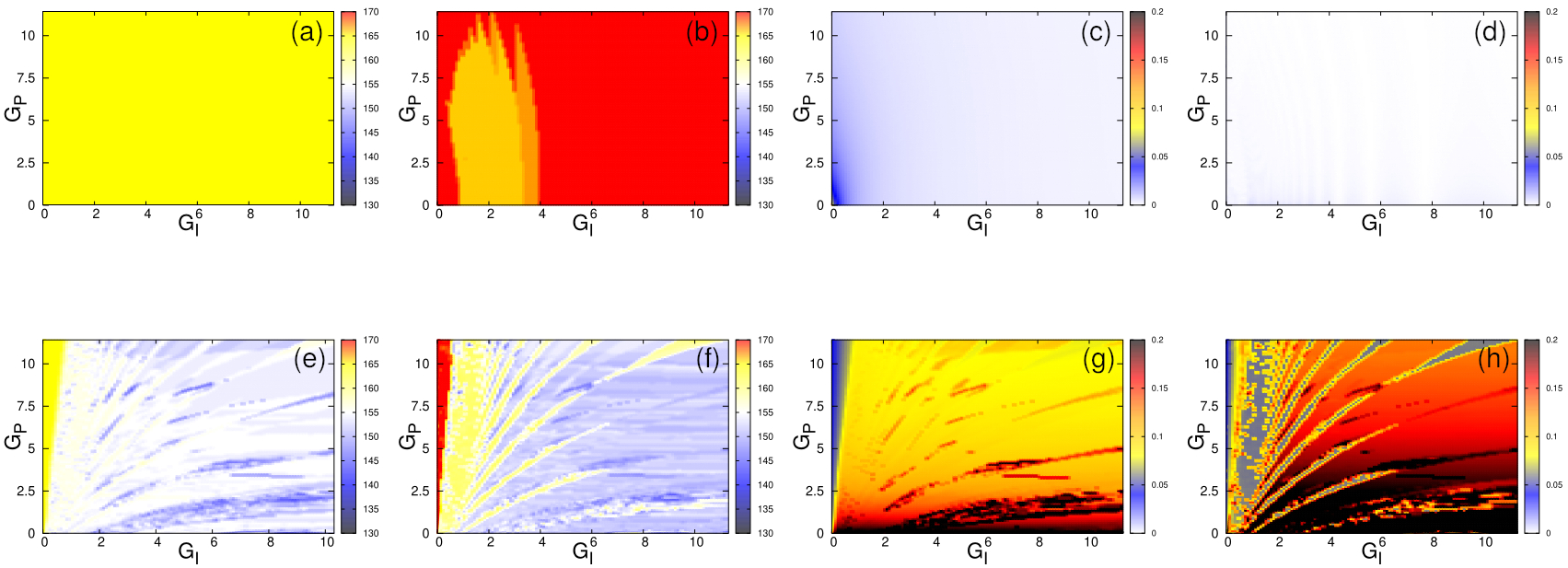}
\caption{Single node perturbation when the proportional layer has the same topology of the physical layer: removal of node 24. First two columns: active links. Third and fourth column: standard deviation $\Delta \omega$ for different values of $G_P$, $G_I$. (a)-(d): Global control topology in the integral layer, i.e., $A^I=A^{ext}(A)$, when the perturbation is active (panels a and c) and when the perturbation is ended (panels b and d). (e)-(h): Local topology in the integral layer, i.e, $A^I=A^{loc}(A)$, when the perturbation is active (panels e and g) and when the perturbation is ended (panels f and h).  All other parameters are set as in Fig. \ref{fig:nodo24extended}. \label{fig:nodo24Italianextended}}
\end{figure*}

We move to discuss the scenario where the topology of the proportional layer is the same of the physical layer. We have found that, under such conditions, controlling the cascading failures becomes more challenging, regardless of the selected integral layer topology. In Fig.~\ref{fig:nodo24Italianextended} we analyse this scenario for different values of $G_P$ and $G_I$, when the integral layer has a global (panels a-d) or local (panels e-h) topology, i.e., $A^I=A^{ext}(A)$ or $A^I=A^{loc}(A)$ respectively. In comparison with the corresponding panels of Fig.~\ref{fig:nodo24extended}, it turns out that using a local topology $A^I=A^{loc}(A)$ yields a lower level of synchronization (panels g, h) and a smaller region where cascading failures are fully controlled (panel f).

\begin{figure}
\centering
\includegraphics[width=0.45\textwidth]{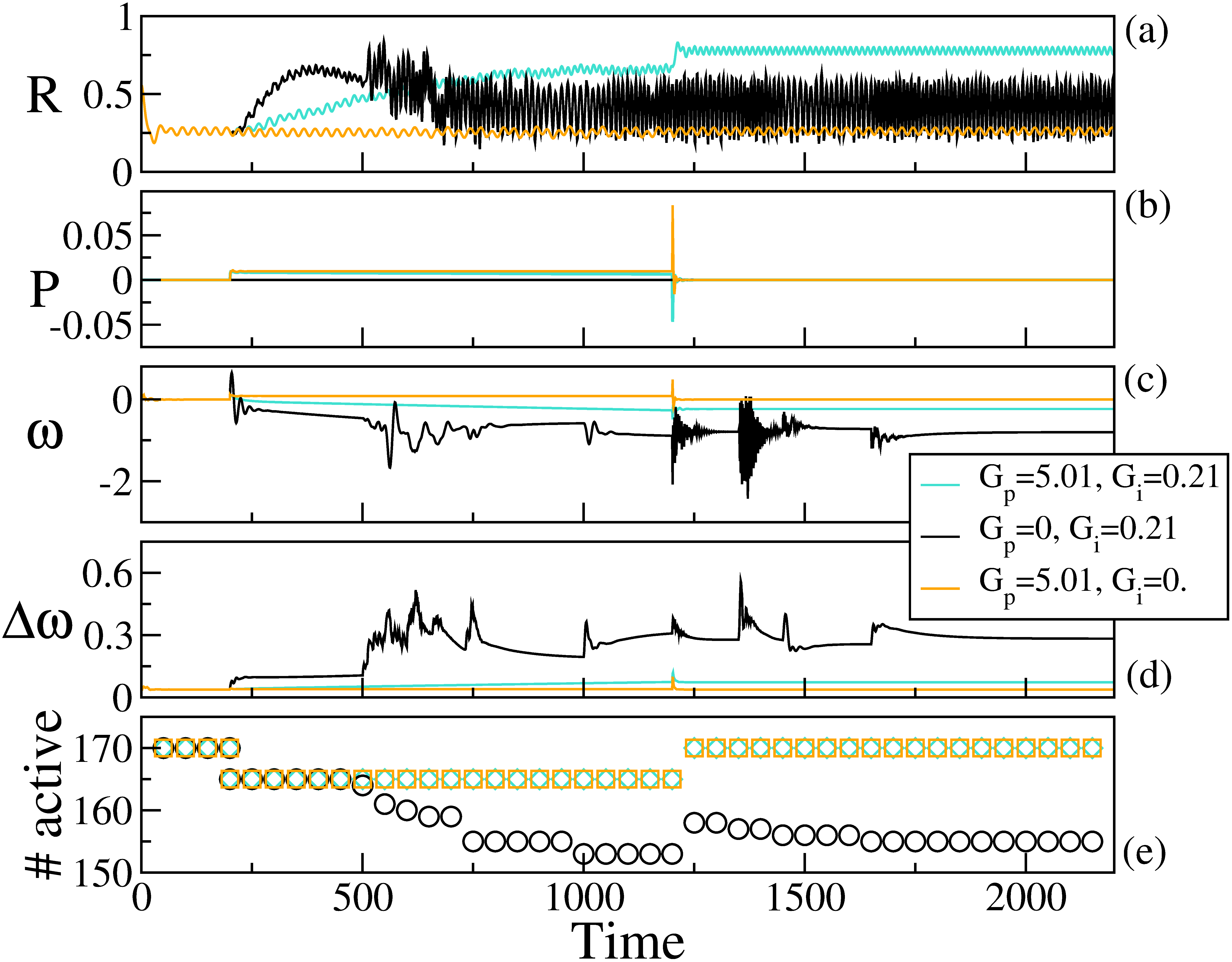}
\caption{Time behavior of the order parameter $R$ (a), the power loss of all loads (b), the frequency $\omega_i$ of the perturbed node (c), the standard deviation $\Delta \omega$ (d), and the number of active links (e) when node 24 is removed. The proportional layer has the same topology of the physical layer. The integral layer is characterized by the adjacency matrix $A^I=A^{loc}(A)$. The other parameters are set as in Fig. \ref{fig:nodo24extended}. 
\label{fig:nodo24timebehaviorItalian}}
\end{figure}

\begin{figure*}
\includegraphics[width=1\textwidth]{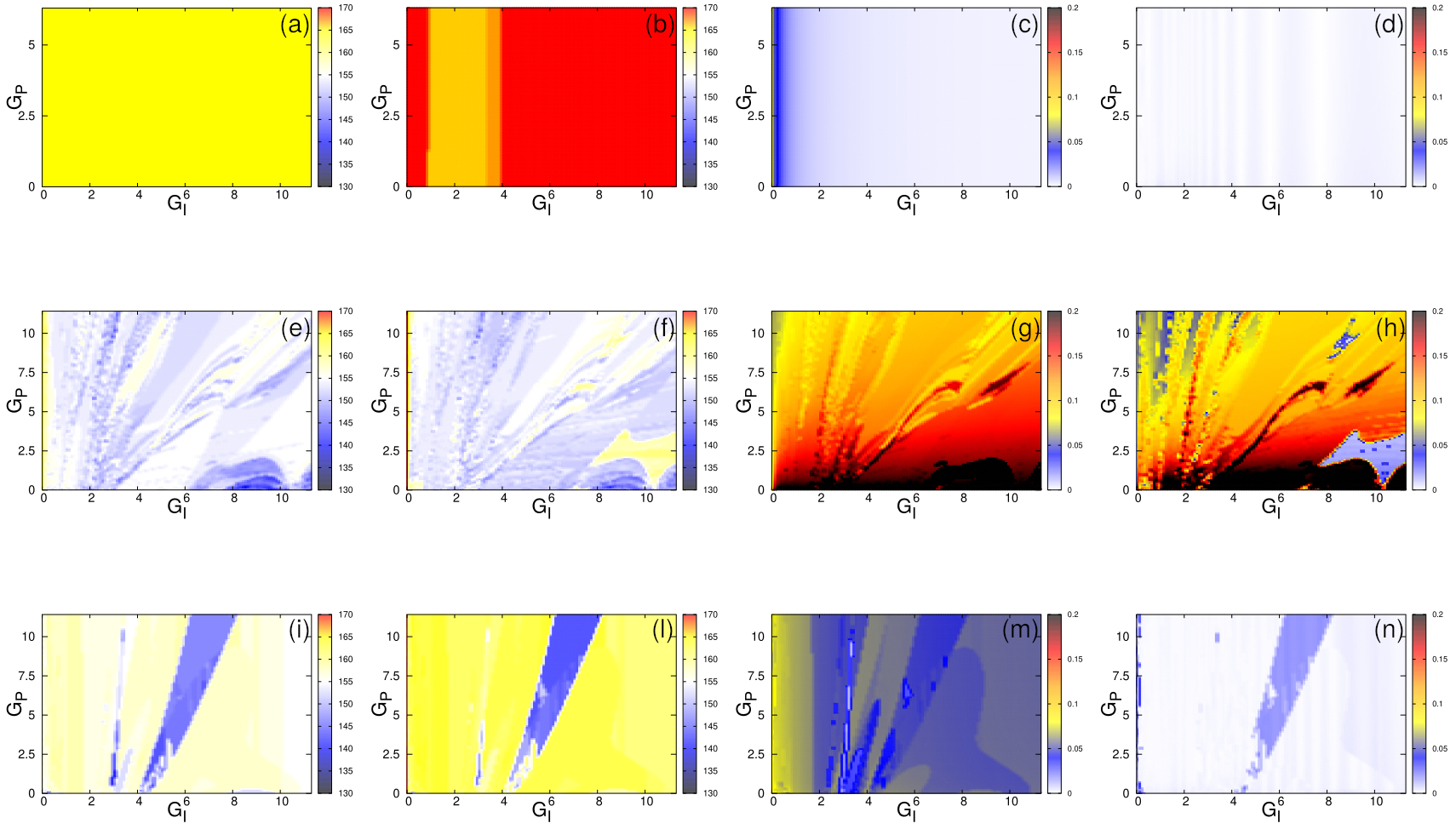}
\caption{Single node perturbation when the proportional layer is given by an ER network with $p=0.04$, in case of pinning control (control acts on generators only): removal of node 24. First two columns: active links. Third and fourth column: standard deviation $\Delta \omega$ for different values of $G_P$, $G_I$. (a)-(d): Global control topology in the integral layer, i.e., $A^I=A^{ext}(A)$, when the perturbation is active (panels a and c) and when the perturbation is ended (panels b and d). (e)-(h): Local topology in the integral layer, i.e, $A^I=A^{loc}(A)$, when the perturbation is active (panels e and g) and when the perturbation is ended (panels f and h).  In both cases the proportional layer has the same topology of the physical layer. All other parameters are set as in Fig. \ref{fig:nodo24extended}. (i)-(n): Local control topology in the integral layer, i.e., $A^I=A^{loc}(A^P)$, when the perturbation is active (panels i and m) and when the perturbation is ended (panels l and n).  Parameters as in Fig.~\ref{fig:nodo24extended}. \label{fig:nodo24Pinning}}
\end{figure*}

\begin{figure}
\centering
\includegraphics[width=0.45\textwidth]{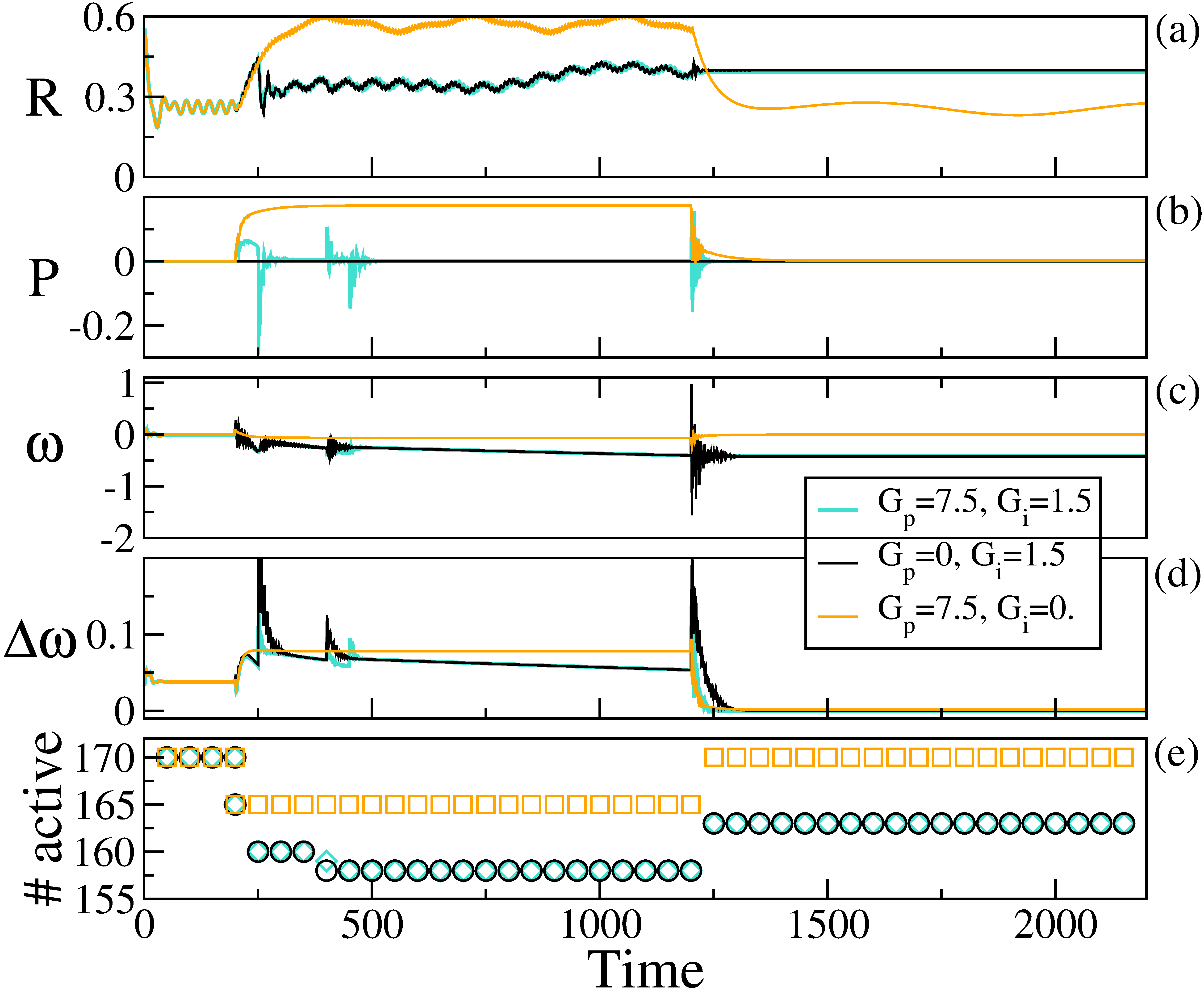}
\caption{
Time behavior of the order parameter $R$ (a), the power loss of all loads (b), the frequency $\omega_i$ of the perturbed node (c), the standard deviation $\Delta \omega$ (d), and the number of active links (e) when node 24 is removed. The proportional layer is an ER random network with connection probability $p=0.04$. The integral layer is characterized by the adjacency matrix $A^I=A^{loc}(A^P)$. In the proportional layer pinning control of only generator nodes is considered. The other parameters are set as in Fig.~\ref{fig:nodo24extended}.
\label{fig:nodo24timebehaviorPinning}}
\end{figure}

\begin{figure*}
\includegraphics[width=1\textwidth]{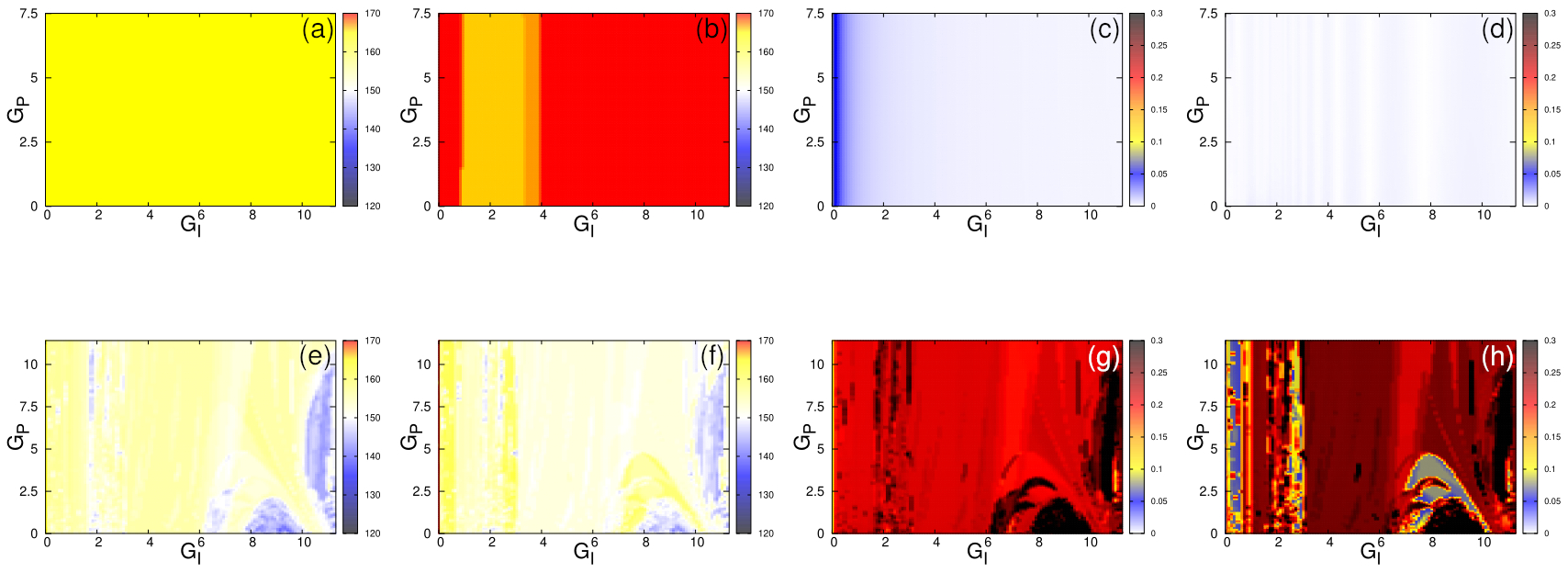}
\caption{Single node perturbation when the proportional layer has the same topology of the physical layer, in case of pinning control (control acts on generators only): removal of node 24. First two columns: active links. Third and fourth column: standard deviation $\Delta \omega$ for different values of $G_P$, $G_I$. (a)-(d): Global control topology in the integral layer, i.e., $A^I=A^{ext}(A)$, when the perturbation is active (panels a and c) and when the perturbation is ended (panels b and d). (e)-(h): Local topology in the integral layer, i.e, $A^I=A^{loc}(A)$, when the perturbation is active (panels e and g) and when the perturbation is ended (panels f and h). The other parameters are set as in Fig.~\ref{fig:nodo24extended}.
\label{fig:nodo24PinningItalian}}
\end{figure*}

\begin{figure*}
\includegraphics[width=1\textwidth]{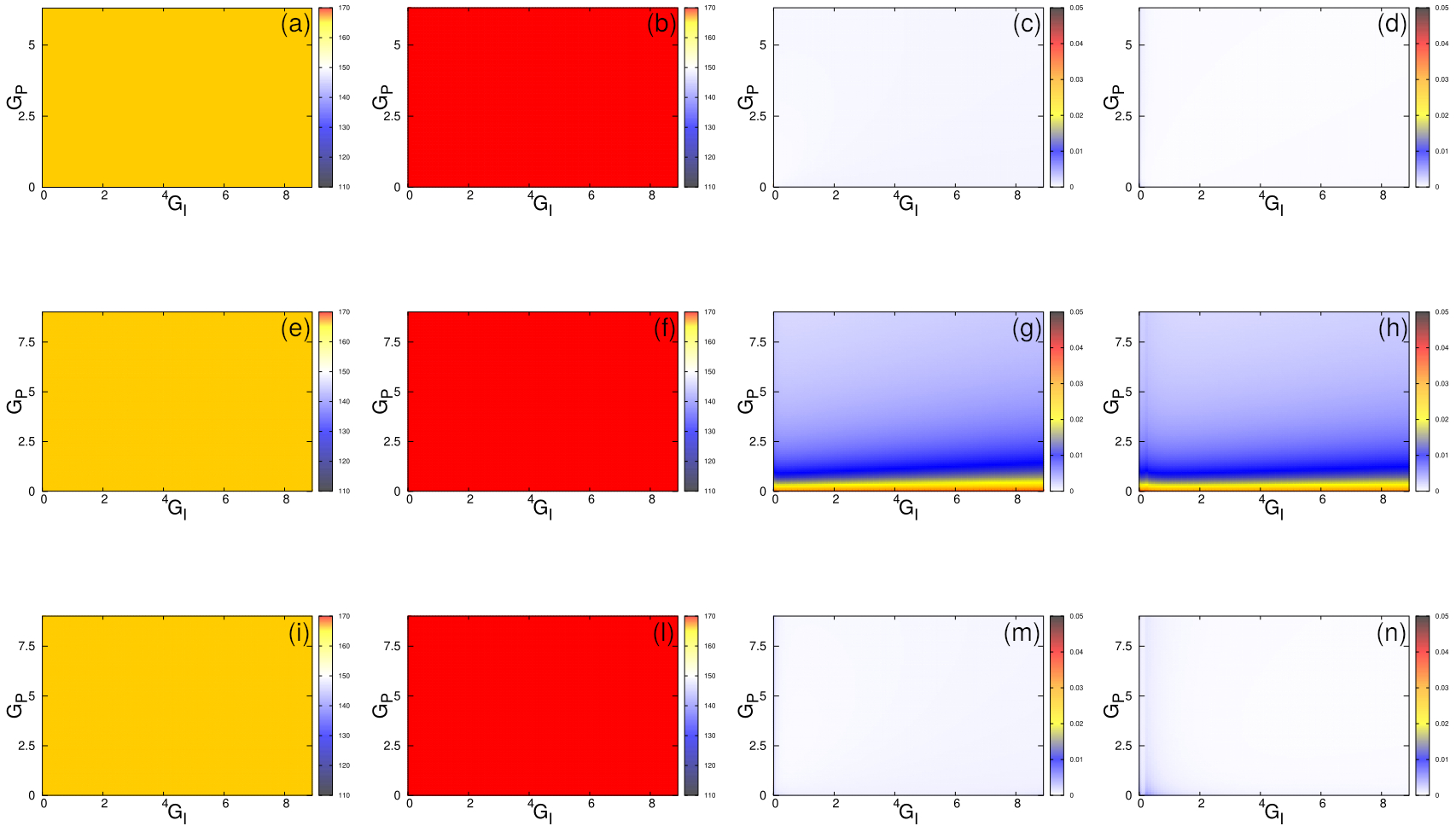}
\caption{Single node perturbation when the proportional layer is given by an ER network with $p=0.04$, in case of pinning control (control acts on generators only): removal of node 28. First two columns: active links. Third and fourth column: standard deviation $\Delta \omega$ for different values of $G_P$, $G_I$. (a)-(d): Global control topology in the integral layer, i.e., $A^I=A^{ext}(A)$, when the perturbation is active (panels a and c) and when the perturbation is ended (panels b and d). (e)-(h): Local topology in the integral layer, i.e, $A^I=A^{loc}(A)$, when the perturbation is active (panels e and g) and when the perturbation is ended (panels f and h).  (i)-(n): Local control topology in the integral layer, i.e., $A^I=A^{loc}(A^P)$, when the perturbation is active (panels i and m) and when the perturbation is ended (panels l and n). The other parameters are set as in Fig.~\ref{fig:nodo24extended}.
\label{fig:nodo28Pinning}}
\end{figure*}

Also for this scenario, we show the time evolution of some quantities of interest in Fig.~\ref{fig:nodo24Italianextended} for selected values of $G_P$ and $G_I$. In particular, we consider an optimal choice of these parameters (panels b and d in Fig.~\ref{fig:nodo24Italianextended}), and compare the results with the cases when only one mode of the controllers is active. In all cases, we observe the emergence of oscillating dynamics in the order parameter $R$ (see Fig.~\ref{fig:nodo24timebehaviorItalian}). The macroscopic oscillations in $R$ suggest the presence of clusters of whirling oscillators \cite{olmi2014hysteretic}, that relegate the system in a partially synchronized state. When only the proportional control is active (orange curves), the level of synchronization is low ($R\approx 0.25$), the power loss is higher (panel b), but cascading failures are prevented (panel e). On the other hand, when only the integral control is active (black curves), during the onset of the perturbation, a cascading failure is observed and, at the same time, the standard deviation $\Delta \omega$ deviates from zero (panel d), thus resulting in chaotic dynamics of $R$ (panel a). When both control systems are active (light blue curves), the best performance in terms of synchronization level (panel a) and number of active lines (panel e) is obtained.

Another important case study arises when the cyber-layers fail together with the physical nodes, i.e. perturbations are extended to the control layer nodes associated to the physical ones. To investigate this case study, we have carried out a series of simulations, performed under the same conditions as those in Figs. \ref{fig:nodo24extended}, \ref{fig:nodo24Italianextended}. In this case, when the physical node is disconnected, the cyber-nodes controlling the perturbed physical node and the related connections are disabled. Numerical simulations reveal an equivalent scenario (results not shown). In more detail, disconnecting the controlling nodes together with the physical node enhances the possibility to control the cascading failures when we implement a local topology in the integral layer ($A^{loc}(A)$ or $A^{loc}(A^P)$), for both considered topologies in the proportional layer.  However, disconnecting the control nodes induces a higher level of desynchronization during the perturbation.

So far, we have considered the situation where, in the proportional control layer, both generators and loads are controlled, while, in the integral control layer, only generators are controlled. We now study the application of pinning control in both control layers. The results are illustrated in  Figs.~\ref{fig:nodo24Pinning},~\ref{fig:nodo24timebehaviorPinning}, and~\ref{fig:nodo24PinningItalian}.  In Figs.~\ref{fig:nodo24Pinning} and~\ref{fig:nodo24timebehaviorPinning} the proportional layer has topology given by the ER network with $p=0.04$, while in Fig.~\ref{fig:nodo24PinningItalian} it has the same topology of the physical layer. Altogether these results show that, when pinning control is applied in the proportional control layer, it is more difficult to prevent cascading failures, regardless of the topology used in the integral control layer. Due to the challenges in preventing cascading failures, maintaining synchronization becomes more difficult as well, as it is particularly evident when considering the local control topology $A^I=A^{loc}(A)$ in the integral layer (see panels g,h in Figs.~\ref{fig:nodo24Pinning} and~\ref{fig:nodo24PinningItalian}). The time evolution of the system dynamics in Fig.~\ref{fig:nodo24timebehaviorPinning} refers to the case where the proportional layer has topology given by the ER network and the integral layer has adjacency matrix $A^I=A^{loc}(A^P)$, and reveals the emergence of oscillating dynamics in the order parameter $R$ (see Fig. \ref{fig:nodo24timebehaviorPinning} a). If both controllers are active (light blue curve) we observe no synchronization during the perturbation and partial synchronization ($R\approx 0.4$) when the perturbation ends, analogously to the case when only the integral control is present (black curve). Conversely, if only the proportional control scheme is active, $R$ (orange curve) shows a higher level of synchronization during the perturbation but a lower level of synchronization when the perturbation ends, with respect to the previous cases. While the presence of the integral control increases the number of cascading failures (panel e, light blue and black curves), the dynamics in the presence of the proportional control alone shows a larger power loss (panel b) but full prevention of cascading failures once the node is reconnected (panel e).


Let us move to illustrate another case study, where now the perturbation is applied to node 28. Similarly to node 24, also this node is a generator, but it is connected with two (rather than three) other power units.
As for node 24, also the removal of node 28 does not give rise to a cascading failure, even in the absence of control, according to the analysis reported in Sec.~\ref{cascading_failures}. In this case, the region where it is possible to achieve both cascading failure and synchronization control is wider with respect to the previous cases (see Fig.~\ref{fig:nodo28Pinning}). In particular, in the entire parameter region no cascading failures are observed at the end of the perturbation, regardless of the topology used in the integral control layer. We notice that, in Fig.~\ref{fig:nodo28Pinning}, we have used pinning control in both proportional and integral layer. Full control is found in a region that extends in the whole parameter space also when all nodes are controlled in the proportional layer. While the standard deviation $\Delta \omega$ remains relatively small in all the parameter region both during the perturbation (panels c, m) and under unperturbed conditions (panels d, n), for the extended topology and for the local topology $A^{loc}(A^P)$, the control of synchronization slightly deteriorates for decreasing values of $G_P$, when we apply the local topology  $A^{loc}(A)$ in the integral layer. Similar results have been also found when the proportional control layer is chosen to be equal to the physical layer.

\begin{figure*}
\includegraphics[width=1\textwidth]{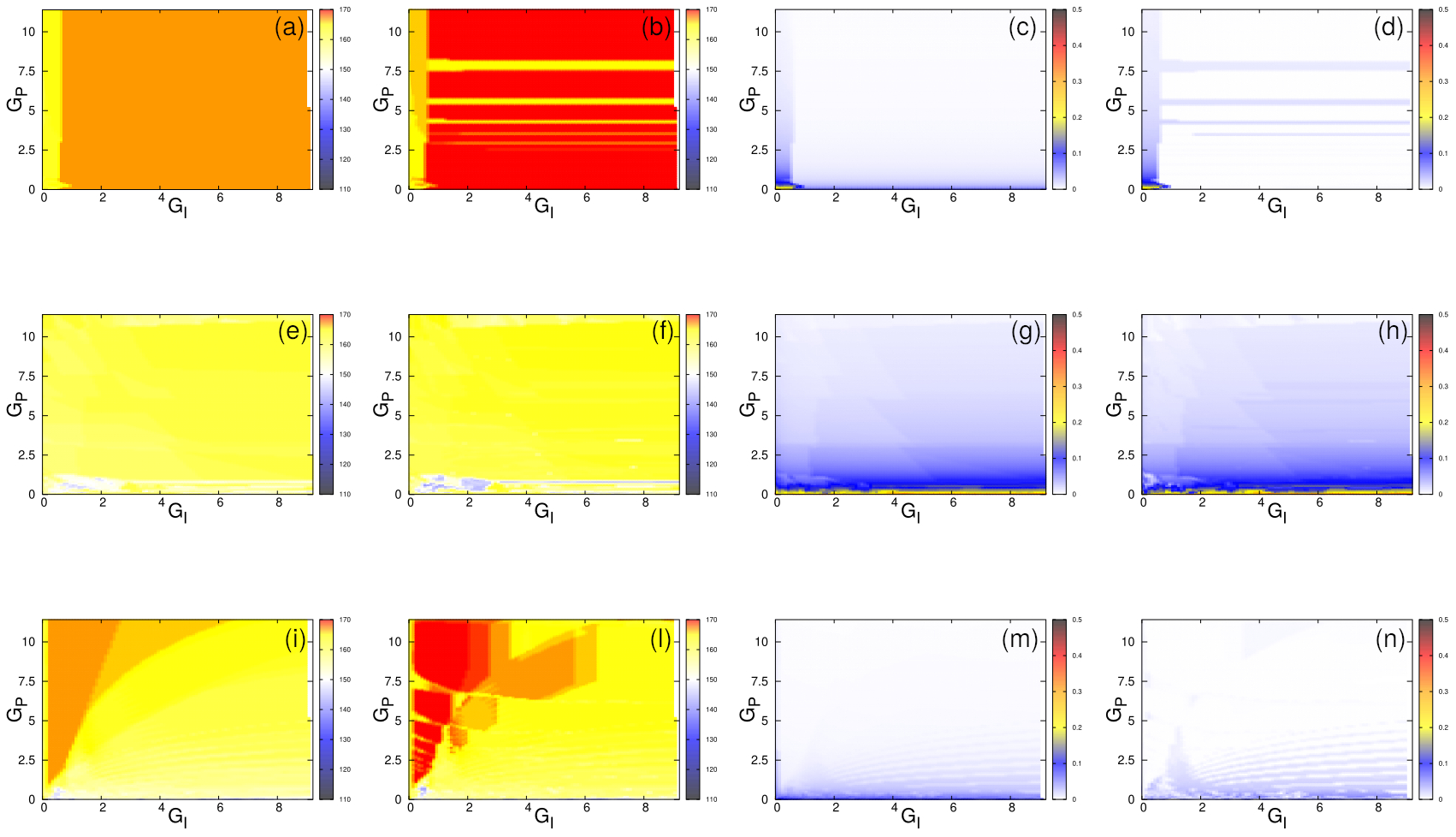}
\caption{Single node perturbation when the proportional layer is given by an ER network with $p=0.04$: removal of node 10. First two columns: active links. Third and fourth column: standard deviation $\Delta \omega$ for different values of $G_P$, $G_I$. (a)-(d): Global control topology in the integral layer, i.e., $A^I=A^{ext}(A)$, when the perturbation is active (panels a and c) and when the perturbation is ended (panels b and d). (e)-(h): Local topology in the integral layer, i.e, $A^I=A^{loc}(A)$, when the perturbation is active (panels e and g) and when the perturbation is ended (panels f and h).  (i)-(n): Local control topology in the integral layer, i.e., $A^I=A^{loc}(A^P)$, when the perturbation is active (panels i and m) and when the perturbation is ended (panels l and n).
The other parameters are set as in Fig.~\ref{fig:nodo24extended}.
\label{fig:nodo10extended}}
\end{figure*}

Lastly, we analyse the effect of the application of the perturbation to node 10, which is a load bus of the power system, connected to two other nodes (both of them are loads). The results for this node are illustrated in Fig.~\ref{fig:nodo10extended}. Node 10 is a critical node, whose disconnection can generate a cascading failure. Since the integral control scheme is designed for controlling generators only, it is more difficult to secure a good synchronization level keeping the standard deviation  $\Delta \omega$ low, as the perturbation is applied, in this case, on a load. Therefore, when the integral layer has a local topology, i.e., $A^I=A^{loc}(A)$, full control of the cascading failures is impossible (panels e, f). This also affects the level of synchronization in terms of standard deviation $\Delta \omega$ that increases when $G_P$ decreases. However, when the integral layer has a local topology induced by the proportional layer, i.e., $A^I=A^{ext}(A)$, we find a quite large region, occurring for high values of $G_P$ (red region in panel b), where the level of synchronization is large (panel d). The region, where cascading failures are completely avoided, increases for increasing $G_P$ and requires a large enough value of $G_P$. The best performance is obtained for $A^I=A^{ext}(A)$, which ensures a wider region where it is possible to simultaneously maintain the system synchronized and prevent cascading failures (panels b and d, respectively)

\section{Conclusions}
\label{sec:conclusions}

In this paper, we have investigated the joint application of a proportional and an integral control layer to a power system subject to large perturbations that can cause the failure of a node of the grid. When this occurs, the system may experience both loss of synchrony and the onset of a cascading failure. To model the power system, we have considered swing equations coupled with an overflow condition that implements a simple shutdown mechanism for the lines. As demonstrated in \cite{fan2022network}, incorporating other protection mechanisms into dynamical models of power grids can crucially result in cascading failures having different sizes and involving different lines. Considering this, our model represents a parsimonious choice in relation to the primary research question of the paper and the associated computational efforts to address it. We expect that our findings remain applicable under diverse assumptions about the power grid dynamics and the protection mechanisms, albeit with potentially varying quantitative outcomes.

Our main result is that the multi-layer approach is effective in maintaining a high level of synchronization, while simultaneously preventing the occurrence of cascading failures. To achieve this, the coupling coefficients in the two control layers, namely the control gains, need to be tuned. Although, from a control perspective, having analytic formulas or even a chart to guide the tuning of these parameters would be desirable, we have found that the non-trivial interplay between topologies of the two layers, dynamical parameters of the model, and the properties of the node perturbed makes difficult to find a general solution to this problem. For this reason, one can resort to the numerical analysis of the system behavior vs. the control gains, which allow for the identification of the region where control is effective, once fixed the system features above mentioned. Another aspect to be critically examined is the possibility of reducing the number of nodes where control is applied. While we have found that the integral control may be applied only to generators, effectively reducing the set of nodes to control, our findings seem to indicate that the same does not hold, in general, for the proportional layer, which requires a larger sets of nodes to control.

\section*{Acknowledgments}

This work was supported by the Italian Ministry for Research and Education (MIUR) through Research Program PRIN 2017 under Grant 2017CWMF93, project "Advanced Network Control of Future Smart Grids - VECTORS".

\section*{Data availability}

All numerical results have been obtained by C or MATLAB code developed by the Authors. The network evolution equations  have been integrated by employing a 4$th$ order Runge-Kutta scheme with fixed time step $dt = 0.01$. The data associated with this study are available from the corresponding author upon reasonable request. 

\section*{Declarations}

\textbf{Conflict of interest.} The authors have no conflicts of interest.

\bibliographystyle{IEEEtran}
\bibliography{miobiblio}
\end{document}